\documentclass[aps,prl,twocolumn,showpacs,superscriptaddress,groupedaddress]{revtex4}
\usepackage[centertags]{amsmath}
\usepackage{amsmath}
\usepackage[dutch,british]{babel}
\usepackage{amsfonts}
\usepackage{amssymb}
\usepackage{amsthm}
\usepackage{newlfont}
\usepackage{color}
\usepackage{subfig}
\usepackage{bbm}

\usepackage{stmaryrd}
\usepackage{mathrsfs}
\usepackage{fancyhdr}
\usepackage{graphicx}
\usepackage{fancybox}
 \usepackage{psfrag}
 
 \usepackage{multirow}

\newcommand{\ket}[1]{|#1\rangle}
\newcommand{\bra}[1]{\langle#1|}

\newcommand{\bbZ}{\mathbb{Z}}
\newcommand{\id}{1}
\newcommand{\tr}{\mathrm{tr}}
\theoremstyle{plain}

\newcommand{\ran}{\mathrm{ran}}

\begin{document} 

\title{A many-body Fredholm index for ground state spaces {and Abelian anyons}}
\author{Sven Bachmann}
\address{The University of British Columbia, Vancouver, BC V6T 1Z2, Canada}
\email{sbach@math.ubc.ca}

\author{Alex Bols}
\address{University of Copenhagen, DK-2100 Copenhagen \O, Denmark}
\email{alex-b@math.ku.de}

\author{Wojciech De Roeck}
\address{KU Leuven,
3001 Leuven, Belgium}
\email{wojciech.deroeck@kuleuven.be}

\author{Martin Fraas}
\address{Virginia Tech, Blacksburg, VA 24061-0123, USA}
\email{fraas@vt.edu}


\date{\today}
\begin{abstract}
We propose a many-body index that extends Fredholm index theory to many-body systems. The index {is defined for any} charge-conserving system with a topologically ordered $p$-dimensional ground state sector. The index is fractional with the denominator given by $p$. In particular, this yields a new short proof of the quantization of the Hall conductance and of Lieb-Schulz-Mattis theorem. 
In the case that the index is non-integer, the argument provides an explicit construction of Wilson loop operators exhibiting a non-trivial braiding and that can be used to create fractionally charged Abelian anyons. 
\end{abstract}

\maketitle

%\nocite{*}

\noindent \textbf{Introduction.} The use of topology to study condensed matter systems is among the most influential developments of late 20th century theoretical physics   \cite{Haldane_Nobel, Laughlin_Nobel}. The first major application of topology appeared in the context of the quantum Hall effect \cite{TKNN, Thouless85, AvronSeilerSimon83} in the early 80', and topological concepts have since been applied systematically to discover and classify \emph{phases of  matter}
\cite{laughlin1983anomalous,read1999beyond, haldane1983fractional,wen1989vacuum,wen1989chiral,moore1991nonabelions,frohlich1995quantum}.
The full classification for independent fermions is well developed, in particular by K-theory \cite{KitaevTable,schnyder2008classification,heinzner2005symmetry}, but a framework of similar scope is lacking for interacting systems, except possibly in 1 dimension where there is a classification of matrix product states \cite{chen2011classification,TurnerFermions,fidkowski2011topological} and cellular automata \cite{Gross2012, cirac2017matrix}.
For non-interacting systems, several topological indices can be formulated as \emph{Fredholm indices} \cite{fredholm1903,ASSIndex, BellissardSchuba} or, equivalently, as transport through a \emph{Thouless pump} \cite{Thouless83}. These formulations have been influential and insightful, in particular for non-translation-invariant systems \cite{ProdanBook}.  For example, the quantum Hall conductance \cite{ASS90}, the $\bbZ_2$-Kane-Mele index \cite{katsura2016,de2015spectral}, and the particle density can be expressed as (integer-valued) Fredholm indices. 

The aim of this letter is to provide an interacting counterpart to this formalism. In a natural sense, it also gives rise to fractional indices and to Abelian anyons. 

\begin{figure}
\begin{center}
\includegraphics[width=0.475\textwidth]{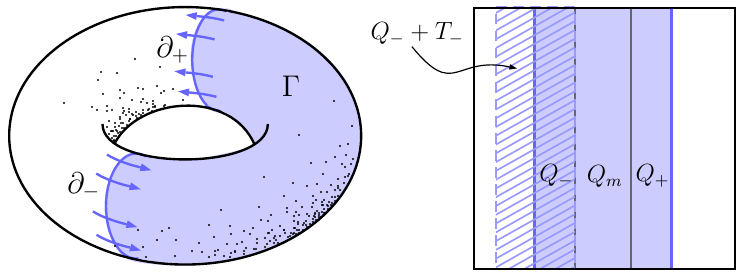}
\caption{The charge transported across the circle $\partial_-$ by the unitary $U$ is exactly compensated by the charge transported across $\partial_+$.}
\label{fig: torus}
\end{center}
\end{figure}

\noindent \textbf{Free fermions.} 
Consider a 2d discrete torus $\mathbb{T}_L$ of $L\times L$ sites $i=(i_1,i_2)$ and let $\Gamma$ be the region $0<i_1\leq L/2$, see Figure~\ref{fig: torus}. 
Let $P$ be an orthogonal projection that we think of as a \emph{Fermi projection} corresponding to a one-particle Hamiltonian on the 2-torus, and let $U$ be a unitary such that $[P,U]=0$. 
These are operators on the (spinless) one-fermion space $\ell^2(\mathbb{T})$. Let $Q$ (charge) be the projector on $\Gamma$:  $Q=1_{\Gamma}=\sum_{i \in  \Gamma} |i\rangle\langle i|$. We consider the charge transported by $U$ out of $\Gamma$ starting from the filled Fermi sea, given by $\tr[ P(U^\dagger QU-Q)]$.  One immediately checks by using $[P,U]=0$ and cyclicity of the trance that this vanishes. This is because the transport at $i_1=0$ is offset by an opposite flow at $i_1=L/2$. Separately however, the flows do not need to be trivial.
If $U$ is sufficiently local, i.e.\ the matrix elements $U({i,j})$ decay fast as $|i-j|\to\infty$, then $U^\dagger QU-Q=(U^\dagger QU-Q)_-+(U^\dagger QU-Q)_+$ with $(U^\dagger QU-Q)_\pm$ located around the boundaries $\partial_\pm$ of $\Gamma$.  Then the charge transport through $\partial_-$ is given by 
\begin{equation}\label{eq: fredholm}
\mathrm{Ind}(P,U)\equiv  \tr[ P(U^\dagger QU-Q)_-].
\end{equation}
If $P$ is also local in the above sense, then $\mathrm{Ind}(P,U)$ is well-defined and it is an integer: $\mathrm{Ind}(P,U) \in \mathbb{Z}$ up to corrections vanishing for large $L$. 
This formula is insensitive to local changes: if we add to any of $Q,P,U$ an operator $B$ well-localized around $\partial_-$, then the index does not change, reflecting its topological nature.  
Our presentation, inspired by  \cite{KitaevToricCode}, was stressing the Thouless pump picture, and we refer to SM for the connection to a Fredholm index and the omitted proof.
In both cases, the point is that the index is constructed in a general way out of the minimal data  provided by $P,U$. In particular, if the index is quantum Hall conductance, its quantization is shown without recourse to any explicit bundle.  

\noindent \textbf{Interacting systems.} 
We consider a many-body setting, either of spins or fermions on the discrete torus $\mathbb{T}_{L}$. 
We say that an observable $O$ has support $X \subset \mathbb{T}_{L}$ \footnote{This is literally true for spin systems. For fermionic systems, an observable $O$ has support $X$ if it is a function of the field operators $c_i,c^\dagger_i, i \in X$} if $O=O_X\otimes \id_{X^c}$.  A \emph{local} observable is supported in a fixed, $L$-independent set $X$, up to rapidly vanishing tails \footnote{The natural notion here is that $O = \sum_{n\geq 0} O_n$, where $O_n$ is supported in the $n$-fattening of $X$ and $\Vert O_n \Vert n^k \to 0$ as $n\to\infty$, for any $k$ and uniformly in $L$.}. All our equalities hold up to finite size corrections of order $\mathcal{O}(L^{-\infty})$, i.e.\ decaying faster than polynomial in $L$, as was also the case above.\\
We  consider a many-body ground state projector $P$ with some finite rank $p$ (dimension of ground state space).  Even though we use the same symbol, this is very different from the Fermi projection above, which is a one-particle concept.   In the interesting case $p>1$, we require the distinct ground states to be locally indistinghuishable, a condition that  is also called \emph{topological order} \cite{wen1990topological,Bravyi:2011ea}
$$
P O P = \mathrm{tr}(P O) \frac{P}{p}
$$
for any local operator $O$.  The charge operator $Q$ is now the number of fermions in $\Gamma$, i.e.\ 
$Q=\sum_{i \in \Gamma} n_i$. This choice is made for the sake of concreteness, the only important feature is that $Q$ is made out of a collection of commuting, local operators with integer spectrum. The operator $U$ is a unitary process that leaves the ground state space invariant $[P,U]=0$ and that conserves the total number of fermions, but of course not necessarily $Q$. 
Therefore $U^\dagger Q U-Q$ is again a sum of two contributions
$T_\pm \equiv (U^\dagger Q U-Q)_{\pm}$ located respectively at $\partial_-,\partial_+$. 
This splitting is in general not uniquely defined and we choose it to satisfy $e^{2\pi i(Q+T_\pm)}=1$, see below for details and an explanation.
Analogously to the free case, we now consider, for any ground state $\psi \in \ran P$, 
\begin{equation}
\label{eq:index}
\mathrm{Ind}(P,U)\equiv \langle \psi| (U^\dagger QU-Q)_-|\psi\rangle.
\end{equation}
The locality that was crucial in the non-interacting setting is now implemented as follows: $1)$ we require the ground projection $P$ to correspond to a  local  Hamiltonian (sum of local terms) $H=\sum_X H_X$ that is gapped, uniformly in volume, and  $2)$ For any operator $O$, the spatial support of $U^\dagger O U$ extends beyond the support of $O$ by a distance that is at most $o(L)$, i.e.\ $\mathrm{distance}/L\to 0$ as $L\to\infty$.\\

\noindent\textbf{Index Theorem.}
\emph{The index $\mathrm{Ind}(P,U)$ is a multiple of $1/p$, i.e.\ $\mathrm{Ind}(P,U) \in \mathbb{Z}/p$.}\\

\noindent The index (\ref{eq:index}) is independent of the choice of $\psi$ in the ground state sector, as follows from topological order since $U^\dagger QU-Q$ is a sum of local terms.
The robustness enjoyed by the noninteracting index \eqref{eq: fredholm} is also present here. For example, if we add to $Q$ an observable $B$ that is a sum of local terms supported around $\partial_-$, the index changes by $\langle \psi|(U^\dagger B U-B)|\psi\rangle$. By topological order and the locality of $B$, the expression takes the same value for any ground state and hence it equals $\tfrac1p \tr P(U^\dagger B U-B)P$. By $[P,U]=0$ and cyclicity of the trace, this vanishes. The index is also additive. If $U_j, j=1,2$ are two unitaries satisfying  the  assumptions with corresponding transported charges $T_\pm^{(j)}$ then
$
U_1^\dagger U_2^\dagger Q U_2 U_1 = Q + T_- + T_+
$
with $T_- = T^{(1)}_- + U_1^\dagger T^{(2)}_-U_1$
and hence we get
\begin{equation}
\label{additivity}
\mathrm{Ind}(P,U) = \mathrm{Ind}(P,U_1) + \mathrm{Ind}(P,U_2).
\end{equation}

Both the non-interacting and the interacting setup can be seen as a Thouless pumps. They construct in a natural way an index out of $P$ and $U$. A significant difference is the possibility of rank $p >1$, which gives rise to an rational index in $\mathbb{Z}/p$.   Related approaches are found in \cite{Oshikawa_Commensurability,Oshikawa_Constraints,matsugatani2018universal,BBDF}.

\noindent \textbf{Splitting.}
As already mentioned, there is a potential ambiguity in the splitting $U^\dagger Q U - Q=T_-+T_+ $. Indeed, if $T_{\pm}$ are valid choices, then so are  $T_\pm \pm j \id$, for any real number $j$.  
There is a canonical physical choice in the case that $U =\mathrm{T} e^{i\int_0^1 ds G(s)}$ (time-ordered exponential) for a family of charge-conserving local Hamiltonians $G(s)$.  Indeed, let $G=G_-+G_m+G_+$ be a splitting of the Hamiltonian $G$ (in charge-conserving terms) according to a partition of $\Gamma$ (see Figure~\ref{fig: torus}), then we can set
$
T_\pm := \mathrm{T} e^{ i\int_0^1 ds [G_{\pm}(s),\cdot]} Q -Q.
$
Because of the commutator and charge conservation, this is independent of the chosen splitting of $G$. Since then $Q+T_\pm$ is unitarily conjugated to $Q$, our condition
$e^{2\pi i(Q+T_\pm)}=1$ is indeed satisfied.  Together with $U$ being translation on the lattice, this case actually covers all interesting examples known to us. 
Let us now argue why the condition $e^{2\pi i(Q+T_\pm)}=1$ can be satisfied in general. We split $Q=Q_-+Q_m+Q_+$ (see Figure~\ref{fig: torus}) so that the three parts commute and have integer spectrum. We  now demand that also $Q_- + T_-$ has integer spectrum (this is equivalent to $e^{2\pi i(Q+T_\pm)}=1$) as it represents the total charge that eventually is present in a neighborhood of $\partial_-$.  Let's prove that such choice exists:  $U^\dagger Q U=(Q_- + T_-) + Q_m + (Q_+ + T_+)$ where the summands have disjoint supports. Since $U^\dagger Q U$ and $Q_m$ have integer spectrum, the spectrum of $(Q_\pm + T_\pm)$ necessarily lies in $\bbZ\pm a$ and we can choose $j$ such that $(Q_\pm + T_\pm)$ has integer spectrum. The remaining freedom $j\in\bbZ$ is harmless to our results.

%\textbf{IN PROGRESS FROM HERE ON}

\subsection{Proof of the index theorem}

\noindent \textbf{Adiabatic Flux Insertion.}
Let us define \cite{Hastings:2004go,Sven}
\begin{equation}
\label{HastingsGenerator}
K:= \int dt W(t)   e^{i t H} i[Q,H]  e^{-i t H}
\end{equation}
with $W$ a real-valued, bounded function satisfying $W(t)=O(|t|^{-\infty})$ and $\widehat{W}(\omega)=\frac{1}{i\omega}$ for all $|\omega|\geq \gamma$, with $\gamma$ the spectral gap of the Hamiltonian. 
The properties of $W$ yield that $[K,P]=[Q,P]$.  By the total charge conservation and locality, we see that
$
[Q,H]=J_-+J_+.
$
with $J_\pm$ localized around $\partial_\pm$.
Altogether, this implies that there are $K_{\pm}$ localized around $\partial_\pm$ such that 
\begin{equation}\label{Qbar}
\bar Q : =Q-K_--K_+
\end{equation}
satisfies
$
[\bar Q,P]=0,
$
and hence also $e^{2 \pi i \bar{Q}}$ commutes with $P$. Since $Q_m$ has integral spectrum, the unitary decomposes as product of unitaries along $\partial_\pm$, $e^{2 \pi i \bar{Q}} = e^{ 2 \pi i \bar{Q}_-}e^{ 2 \pi i \bar{Q}_+}$ with $\bar{Q}_\pm = Q_\pm - K_\pm$. By the `Locality lemma' below (we thank Filippo Santi for pointing out an error in the proof in the published version of this article), each of these unitaries alone leaves $P$ invariant. In SM we further explain that $e^{ 2 \pi i \bar{Q}_-}$ is the `quasi-adiabatic' \cite{hastings2010locality} implementation of $2 \pi$ flux threading through $\partial_-$, provided that the Hamiltonian remains gapped during this process.\\

\noindent \textbf{Locality Lemma.} Let $V=V_-V_+$ with $V_{\pm}$ unitaries supported around $\partial_{\pm}$. Then $[P,V]=0$ implies $[P,V_{\pm}]=0$.  \emph{Proof}: By exponential clustering~\cite{HastingsClustering,BrunoClustering},
$
P V \psi =  P V_- P V_+ \psi
$
for any ground state $\psi \in \ran P$. Then, on one hand $\|PV\psi\| =1$ by the assumption $[P,V]=0$ , on the other hand
$
\Vert P V_+ P V_- \psi\Vert\leq \Vert V_+P\Vert\,\Vert PV_-\psi \Vert = \Vert PV_-\psi\Vert\leq 1.
$
Hence
$
\Vert PV_-\psi\Vert  = 1,
$
and since $\Vert PV_-^\dagger \psi\Vert  = 1$ by the same argument, we conclude that $[P, V_-] = 0$.\\

\noindent \textbf{Core argument.}
We consider
\begin{equation}
\label{D1}
Z_- \equiv U^\dagger e^{2\pi i \bar{Q}_- } U e^{-2 \pi i \bar{Q}_-},
\end{equation}
which will reveal the non-commutativity of $U$ and flux insertion $e^{2\pi i \bar{Q}_- }$. By the locality of $U$, $Z_-$ is supported around $\partial_-$. We are going to show that 
\begin{equation}
\label{eq}
P Z_- P = P e^{\frac{2\pi i}{p}\tr(P T_-)}.
\end{equation}
Since the RHS of~(\ref{D1}) is a product of 4 unitaries commuting with $P$, we have that $\det(P Z_- P) =1$, and (\ref{eq}) then implies $\tr(P T_-) \in \mathbb{Z}$.  The proof is now concluded since, as noted before, the topological order condition implies that for any ground state $\psi$,
$
\bra{\psi} T_- \ket{\psi} = \frac{1}{p} \tr(P T_-).
$
\noindent \textbf{Proof of \eqref{eq}.} By integrality of $Q_m+Q_+$, we can replace $\bar{Q}_-$ by $Q-K_-$ in the first exponential of \eqref{D1}. Bringing $U^\dagger (\cdot) U$ inside the exponential, we write 
$$
U^\dagger(Q-K_-)U = (Q_-+T_--K_-^U) +Q_m +(Q_++  T_+)
$$
where we use a notation $O^U = U^\dagger O U$ and the three bracketed terms commute, see again Figure~\ref{fig: torus}. The exponential of the second/third term is $1$ by integrality/our constraint $e^{2\pi i(Q+T_\pm)} =1$. The $\exp$ of the first term leads to the identity
 $Z_- = e^{2 \pi i (Q_- + T_- - K_-^U)}e^{- 2\pi i \bar{Q}_-}$. 
 We now interpolate between $\id$ and $Z_-$ by the operator
 $
Z_-(\phi) = e^{i \phi ({Q_-} + T_- - K_-^U)}e^{- i \phi \bar{Q}_-},
$
and we prove that $[Z_-(\phi),P]=0$ for all $\phi$.  Indeed, let us introduce the corresponding anti-twist 
$
Z_+(\phi) \equiv e^{i \phi ({Q_+} - T_+ - K_+^U)}e^{- i \phi \bar{Q}_+}
$
Then we see that $Z_-(\phi) Z_+(\phi) = U^\dagger e^{i \phi \bar{Q}} U e^{- i \phi \bar{Q}}\equiv Z(\phi)$ because far from $\partial_\pm$, the charge is unaffected by $U$.   By $[\bar{Q},P]=0$, $Z(\phi)$ commutes with $P$ and hence, by the Locality Lemma above, so do both $Z_\pm(\phi)$ as claimed.

We now differentiate $Z_-(\phi)$ w.r.t.~$\phi$,
\begin{equation} \label{eq: derivative}
\partial_\phi (P Z_-(\phi) P) = P Z_-(\phi) e^{i \phi \bar{Q}_-} i(T_- - K_-^U + K_- )e^{- i \phi\bar{Q}_-} P.
\end{equation}
The quantity in $(\ldots)$, which we name $D_-$ is localized around $\partial_-$ so we can replace $e^{ i \phi\bar{Q}_-}D_-e^{- i \phi\bar{Q}_-}$ by  $e^{i \phi\bar{Q}}D_-e^{- i \phi\bar{Q}}$ and subsequently commute $e^{- i \phi\bar{Q}}$ with $P$. Using also $[Z_-(\phi),P]=0$, we then rewrite \eqref{eq: derivative} as 
$$
\partial_\phi (P Z_-(\phi) P) =  i P Z_-(\phi) P  e^{i \phi \bar{Q}}  P  D_-P e^{- i \phi\bar{Q}} .
$$
We now note that $PD_-P = PT_-P$ since $[U,P]=0$. Furthermore, $T_-$ is a sum of local terms and hence topological order yields $PT_-P=P \frac{1}{p}\tr (PT_-)$. This means that we can also drop the factors $ e^{\pm i \phi \bar{Q}}$, because  of $ [P,\bar{Q}]=0$.  We hence end up with a simple differential equation whose solution, evaluated at $\phi=2\pi$, is \eqref{eq}.

\subsection{Examples}
We focus on applications of the index theorem to systems with degenerate ground state manifold.% Many examples in systems with a unique ground state are given in \cite{BBDF}. 

\noindent {\bf Fractional Lieb-Schultz-Mattis theorem.} Let $U$ be spatial translation by a one site to the left. Then $T_- = Q_{\{x_1 = -1\}}$ is the charge operator in the hyperplane $\{x_1 = -1\}$. If the Hamiltonian is translation-invariant then $\rho = \bra{\psi} T_- \ket{\psi}$ is the charge in any plane $\{x_1=k\}$ and the theorem implies that 
$
\rho \in \mathbb{Z}/p.
$
Of course, $\rho$ should scale $\propto L$ but the result is still meaningful as the equality holds up to $\mathcal{O}(L^{-\infty})$.
This theorem was already basically contained in the original treatments \cite{LSM,Oshikawa_Commensurability,HastingsLSM,BrunoLSM}. 

\noindent {\bf Quantization of Hall conductance.} 
Here we rename $U_1\equiv U=e^{2\pi i \bar{Q}_-}$ and we let $U_2$ be the analogous operator with $\partial_-$ being replaced by the orthogonal loop $\{x_2 = -1/2\}$, i.e. $U_2$ is constructed as $U_1$ upon replacing $Q$ by $\sum_{i  \in \Gamma_2 } n_i$ with $\Gamma_2=\{0<i_2\leq L/2\}$. Now $T_-$ is simply the charge transported by threading a unit of flux in the $2$-direction. This equals the Hall conductance $\sigma$ by the well-known Laughlin argument~\cite{Oshikawa_Constraints,BBDF}. Putting back physical units, our result is that  
$$
\sigma = \frac{q}{p} \frac{e^2}{h}.
$$
 This gives a mathematically rigorous proof of fractional quantization of $\sigma$ in an interacting setting that is shorter than previous arguments in \cite{HastingsMichalakis,giuliani2017universality, BBDF0}.

\noindent {\bf Fractional Avron-Dana-Zak relations.} A fractional quantum Hall sample pierced by a rational flux $\phi$ has a Hamiltonian that is invariant under magnetic translations, which is a composition of a translation and threading the torus by $-\phi$ flux. Combining the discussions of FQHE and Lieb-Schultz-Mattis theorem, and relying on the additivity property~(\ref{additivity}) of our index,  we get the constraint
$
\rho - \phi \sigma \in \mathbb{Z}/p
$. This relation was derived in \cite{AvronDanaZak} for non-interacting systems (hence $p=1$) and in \cite{Watanabe, Oshikawa_Constraints} for interacting systems.

\subsection{Braiding relations and Abelian anyons}
\begin{figure}
\includegraphics[width=0.3\textwidth]{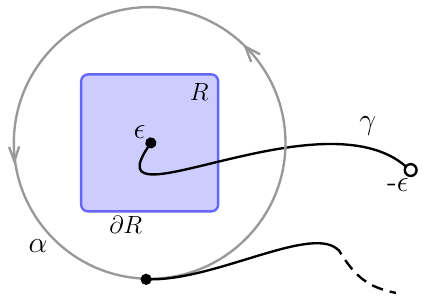}
\caption{The unitary $U_\gamma$ creates a pair of anyonic excitations of opposite charge at the endpoints of $\gamma$.}
\label{fig:braiding}
\end{figure}

Let $U_1,U_2$ be as above in the example of the FQHE. That is, they correspond to threading a unit of flux in the $1,2$-direction. Then, the four unitaries in \eqref{D1} satisfy, by \eqref{eq},
\begin{equation}
\label{commutator}
U_2^\dagger U_1 U_2 U_1^\dagger P = e^{2 \pi i \frac{q}{p}} P,
\end{equation}
and we recall that each of them remains unitary when restricted to $\ran P$. Note that these restricted unitaries  are naturally associated to oriented loops winding around the torus.  If $\tfrac{q}{p}$ is noninteger, then \eqref{commutator} gives a nontrivial commutation relation between those loops, see~\cite{leinaas1977theory,nayak2008non,arovas1985statistical}.    In the case  when $p>1$ and $q,p$ are coprime, and the \emph{topological quantum field theory} (TQFT) describing the ground state sector $\ran P$ is a $U(1)$-Chern Simons theory, these loops can be identified with \emph{Wilson loops}. In particular, the action of $PU_1,PU_2$ on any ground state $\psi$ generates the full ground state sector. This follows because there is no representation of~\eqref{commutator} on a space of dimension smaller than $p$.
As far as we know, our approach is the first explicit construction of such loop operators in generic microscopic models, cf.\ \cite{KitaevToricCode,levin2005string}. \\
\noindent \textbf{Anyonic quasiparticles.}
To any region $\Omega$, we associate $\bar Q_\Omega = Q_\Omega - K_{\partial\Omega}$, where the notation is a reminder of the fact that $K_{\partial\Omega}$,   defined as in \eqref{HastingsGenerator} with $Q\to Q_\Omega$, is an operator supported on the boundary of the domain. 
 By the integrality of the spectrum of $Q_\Omega$, $U_{\partial\Omega} = e^{2\pi i \bar Q_\Omega}$ is a loop operator supported around $\partial\Omega$. 
We can write it explicitly as $U_{\partial\Omega} = \mathrm{T}e^{-i\int_0^{2\pi} d\phi K_{\partial\Omega}(\phi)  }$ where $K_{\partial\Omega}(\phi)= e^{-i\phi Q_\Omega} K_{\partial\Omega}e^{i\phi Q_\Omega}$. Since $K_{\partial\Omega}$ is a sum of local terms, we can choose, albeit not in any canonical way, to retain only the terms associated to an open string  $\gamma\subset\partial\Omega$ and this defines $U_{\gamma}$. Since all local terms in $K_{\partial\Omega}$ commute with the total charge, so does $U_{\gamma}$. 
For a ground state $\psi$, $\varphi = U_{\gamma} \psi$ is a state with two localized excitations at the endpoints of $\gamma$, see Figure~\ref{fig:braiding}.
Indeed, $\varphi$ and $\psi$ are locally indistinguishable away from the endpoints of $\gamma$.  The charge of an excitation is the excess charge in a region~$R$ around the excitation that does not extend to the other endpoint. It is given by
 $$
\epsilon = \langle \varphi | Q_R |\varphi \rangle - \langle \psi | Q_R | \psi \rangle = \langle \psi, (U_\gamma^\dagger Q_R U_\gamma - Q_R) \psi \rangle.
 $$
By charge conservation, $U_\gamma^\dagger Q_R U_\gamma - Q_R$ is supported at the intersection $\partial R\cap\gamma$, so that the excitation has a fractional charge
$$
\epsilon = \frac{q}{p}
$$
by applying the index theorem. The excitation at the other end point has opposite charge.

The factor $q/p$ also appears when braiding the excitations. For a closed contractible path $\alpha$, $U_\alpha \psi$ is proportional to $\psi$, and we set the phase to be $0$. When an excitation is present inside~$\alpha$, the loop is not contractible anymore, and we obtain by  (\ref{commutator})
$$
U_\alpha \varphi = U_\gamma ( U_\gamma^\dagger U_\alpha U_\gamma U_\alpha^\dagger) \psi = e^{2\pi i \frac{q}{p}}\varphi.
$$
Hence, the created excitations are Abelian anyons.

\subsection{Conclusions}
We described an index for systems with $U(1)$ symmetry (charge conservation), reminiscent of the Fredholm index.  The index is associated to a charge transported across a hypersurface and it is rational, with denominator $p$ being the dimension of a topologically ordered ground state sector. 
 We relate the index to a commutation relation on the ground state space, and show that the relation reveals the existence of anyonic excitations whenever the index is non-integer.

\begin{acknowledgments}
\subsection{Acknowledgments}
M.F. was supported in part by the NSF under grant DMS-1907435. The work of S.B. was supported by NSERC of Canada. W.D.R. acknowledges the support of the Flemish Research Fund FWO undergrants G098919N and G076216N, and the support of KULeuven University under internal grantC14/16/062. A.B. was supported by VILLUM FONDEN through the QMATH Centre of Excellence (grant no. 10059).
\end{acknowledgments}

\bibliographystyle{plain}
 \bibliography{PRL_Index}
 
 \newpage
 \clearpage

\section{Supplementary Material}

Firstly, we present the proof, inspired by~\cite{KitaevHoneycomb}, of the index theorem for free fermions. Secondly, we give an explicit expression for the unitary associated with the process of quasi-adiabatic flux insertion. Although the expression is new, all its properties are well-known, see~\cite{HastingsMichalakis, BBDF0}.

\section{Index theorem for free fermions}
We briefly review the setup. 
Let $\mathbb{T}=\mathbb{T}_L$ be the $L \times L$ discrete torus. We say that an operator $O$ on $l^2(\mathbb{T})$ has uniform rapid decay if 
$$
\sup_{i,j, |i-j|\geq \ell} |O_{ij}|  = O(\ell^{-\infty}),
$$
where $|\cdot|$ under the $\sup$ is the graph distance on $\mathbb{T}$. 
A restriction of the operator to a region $\Omega$ is given by $\Pi_\Omega A \Pi_\Omega$ with $\Pi_\Omega = \sum_{i \in \Omega} | i \rangle \langle i|$, and $A\mapsto A_\pm$ is the restriction to a region of width $l$ around $\partial_\pm$, where $\ell \to \infty$ but $\ell/L\to 0$. 

Let $P=P^\dagger=P^2$ be a projection and $U$ a unitary such that both have rapid decay, in the sense above, and such that $[P,U]=O(L^{-\infty})$. We define
\begin{equation*}
\mathrm{Ind}(P,U)=  \tr[ P(U^\dagger QU-Q)_-].
\end{equation*}
The precise statement of the result in the main text is that 
\begin{equation} \label{SM_main}
\mathrm{dist}(\mathrm{Ind}(P,U),\mathbb{Z}) = O(L^{-\infty}).
\end{equation}
We remark that with the conditions given so far we cannot conclude that $\mathrm{Ind}(P,U)$ converges to a fixed integer, as $L\to \infty$, because we did not demand that $P,U$ converge in any way. This could of course easily be done, but it would distract from the main point. 
We now prove \eqref{SM_main} in an approach pioneered by~\cite{KitaevHoneycomb}.

We revert to the convention used in the main text that equalities hold up to $O(L^{-\infty})$ corrections and that $Q = \Pi_\Gamma$ is the charge of the region $\Gamma$ with boundaries $\partial_\pm$.
By rapid decay of $P$, 
$$K =  PQ(1-P) + (1-P) Q P$$
is of the form $K=K_-+K_+$, i.e.\ supported only at $\partial_-  \cup  \partial_+$.  The 
operator
$$
\bar{Q} = Q -K_--K_+, 
$$
commutes with $P$. By rapid decay of $U$ and $[P,U] = 0$, we also have that
$$
U^\dagger \bar{Q} U = Q + (U^\dagger QU-Q)_- + (U^\dagger QU-Q)_+ - K_-^U - K_+^U
$$
commutes with $P$. Here we have again used the shorthand $O^U= U^\dagger O U$.
The two operators $\bar{Q}$ and $U^\dagger \bar QU $ hence commute with $P$. On the other hand, their commutator with $P$ can naturally be decomposed into two terms supported at $\partial_\pm$. These two terms hence have to vanish independently. We conclude that the operator
$$
N = Q + (U^\dagger QU-Q)_- - K_-^U - K_+
$$
also commutes with $P$. Next, we consider the expression 
$$Z_-=U^\dagger e^{2 \pi i \bar{Q}_-} U e^{-2 \pi i \bar{Q}_-}$$ with $\bar{Q}_- = Q -K_-$. We note that by rapid decay $[e^{2 \pi i \bar{Q}_-}, P] = 0$.   Let $\det_P(A) = \det(PAP + (1-P))$. Then, since  $Z_-$ is a product of four unitaries commuting with $P$, we have
$$
\mathrm{det}_P(U^\dagger e^{2 \pi i \bar{Q}_-} U e^{-2 \pi i \bar{Q}_-}) = 1
$$
by the product rule for determinants. On the other hand we have
\begin{align*}
U^\dagger e^{2 \pi i \bar{Q}_-} U e^{-2 \pi i \bar{Q}_-}  &= e^{2 \pi i (Q - K_-^U + (U^\dagger QU-Q)_-)} e^{-2 \pi i \bar{Q}_-} \\
										&= e^{2 \pi i N} e^{-2 \pi i \bar Q}.
\end{align*}
Since both operators in the exponentials commute with $P$, we have
$$
\mathrm{det}_P(U^\dagger e^{2 \pi i \bar{Q}_-} U e^{-2 \pi i \bar{Q}_-}) = e^{2 \pi i (\tr( P(N - \bar Q))}
$$
by the relation between determinant and trace. Plugging the definition of $N$ and using $\tr(P K_-^U )=\tr(PK_-)$ by $[P,U]=0$, this exponential equals $e^{2 \pi i (\tr[ P(U^\dagger QU-Q)_-])}$.
It follows that $\tr[ P(U^\dagger QU-Q)_-]$ is an integer, as was to be proven. 

\section{Adiabatic Flux Threading}

This section refers to the interacting many-body setup. Therefore, the symbols $P,Q,U$ have now a different meaning than the ones in the previous section. 
We use a unitary modelling adiabatic flux threading through the loop $\partial_-$. Let $H(\phi) = e^{i \phi Q} H e^{-i \phi Q}$ be a gauge equivalent `twist-antitwist' Hamiltonian corresponding to threading flux $\phi$ through $\partial_-$ and removing it at $\partial_+$. The ground state projection is then $P(\phi) = e^{i \phi Q} P e^{-i \phi Q}$ and the adiabatic evolution is generated by $Q$. Following~\cite{HastingsWen}, an alternative `quasi-adiabatic' generator  $K(\phi)$ was constructed in \cite{Sven}
\begin{equation}
\label{HastingsGenerator}
K(\phi)= \int dt W(t)   e^{i t H(\phi)} \partial_\phi H(\phi)  e^{-i t H(\phi)},
\end{equation}
with $W$ a real-valued, bounded, integrable function satisfying $W(t)=O(|t|^{-\infty})$ and $\widehat{W}(\omega)=\frac{1}{i \omega}$ for all $|\omega|\geq \gamma$, with $\gamma$ the spectral gap of the Hamiltonian. It satisfies 
\begin{equation*}
\partial_\phi P(\phi) = i [K(\phi), P(\phi)].
\end{equation*}
The advantage of the quasi-adiabatic generator is that it is manifestly supported only in those regions of space where the Hamiltonian actually changes. For the present, charge conserving Hamiltonian, this means that $K(\phi) = K_-(\phi) + K_+(\phi)$, with $K_{\pm}(\phi)$ localized around the loops $\partial_{\pm}$. Furthermore, it satisfies
 $K(\phi) = e^{i \phi Q} K e^{-i \phi Q}$ (we write $K = K(0)$) and from this it follows that the unitary
\begin{equation}
\label{unitaryV}
V(\phi) = e^{i \phi (Q - K)}e^{-i \phi Q} 
\end{equation}
implements the ground state evolution: $P(\phi) = V(\phi)^\dagger P V(\phi)$. 

Of course, the physically more interesting deformed Hamiltonian is one where the flux through $\partial_-$ is not removed at $\partial_+$. It is denoted by  $H_-(\phi)$ and defined \cite{HastingsMichalakis, BBDF0} to be equal to $H(\phi)$ around $\partial_-$ and to $H$ otherwise.  Unlike $H(\phi)$, it is not unitarily equivalent to $H$. If the gap remains open for $H_-(\phi)$ then (\ref{HastingsGenerator}) with $H$ replaced by $H_-$ is the quasi-adiabatic generator associated to $H_-(\phi)$ and by locality it is equal to $K_-(\phi)$.  It follows that the ground state projection $P_-(\phi)$ of $H_-(\phi)$ is obtained by replacing $K \to K_-$ in (\ref{unitaryV}), i.e.
\begin{equation*}
P_-(\phi) = V_-(\phi) P V_-(\phi)^\dagger,\qquad  V_-(\phi) = e^{i \phi(Q - K_-)}e^{-i \phi Q}.
\end{equation*}
In this case and by integrality of the spectrum of $Q$, $e^{ 2 \pi i (Q - K_-)}$ corresponds to a $2 \pi$ flux insertion across $\partial_-$, leaving the GS invariant:
\begin{equation}
\label{2piFlux}
[e^{ 2 \pi i \bar{Q}_-}, P] = O(L^{-\infty}), \qquad  \bar{Q}_- = Q - K_-.
\end{equation}
A remarkable fact \cite{HastingsMichalakis} is that \eqref{2piFlux} holds even if the gap closes at some $\phi \neq 0$.

\end{document}